\documentclass[11pt,dvips]{article}

\usepackage{jheppub}

\usepackage{slashbox,latexsym,amssymb,amsfonts,amsmath,slashed}
\usepackage{graphicx}
\newcommand{\f}[2]{\frac{#1}{#2}}

\newcommand{\la}{\langle}

\newcommand{\ra}{\rangle}

\newcommand{\nyp}[3]{{\bf #1} (#3) #2}
\newcommand{\JHEP}{Jour.\ High Energy Phys.\ }
\renewcommand{\Re}{{\rm Re}\,}

\newcommand{\tr}{{\rm tr}\,}

\title{Deconfinement, chiral transition and localisation in a QCD-like
  model} 

\author[a,b]{Matteo Giordano}
\author[a,b]{\!\!, S\'andor D.\ Katz}
\author[c]{\!\!, Tam\'as G.\ Kov\'acs}%
\author[d]{\!\!, and Ferenc Pittler}%

\affiliation[a]{
Institute for Theoretical Physics, E\"otv\"os University,\\
P\'azm\'any P. s\'et\'any 1/A, H-1117 Budapest, Hungary}

\affiliation[b]{
  MTA-ELTE ``Lend\"ulet'' Lattice Gauge Theory Research Group,\\ P\'azm\'any
  P. s\'et\'any 1/A, H-1117 Budapest, Hungary}

\affiliation[c]{
Institute for Nuclear Research of the Hungarian Academy of Sciences, \\
Bem t\'er 18/c, H-4026 Debrecen, Hungary 
}%

\affiliation[d]{
  HISKP(Theory), University of Bonn,\\
  Nussallee 14-16, D-53115,Bonn, Germany
}

\emailAdd{giordano@bodri.elte.hu}
\emailAdd{katz@bodri.elte.hu}
\emailAdd{kgt@atomki.mta.hu} 
\emailAdd{pittler@hiskp.uni-bonn.de}

\abstract{We study the problems of deconfinement, chiral symmetry
  restoration and localisation of the low Dirac eigenmodes in a toy
  model of QCD, namely unimproved staggered fermions on lattices of
  temporal extension $N_T=4$. This model displays a genuine
  deconfining and chirally-restoring first-order phase transition at
  some critical value of the gauge coupling. 
  Our results indicate that the onset of localisation of the lowest
  Dirac eigenmodes takes place at the same critical coupling 
  where the system undergoes the first-order phase transition. This
  provides further evidence of the close relation between
  deconfinement, chiral symmetry restoration and localisation
  of the low modes of the Dirac operator on the lattice.
}


\begin{document}
\maketitle

\section{Introduction}
\label{sec:intro}

The breaking of chiral symmetry at low temperatures is a crucial
phenomenon in QCD, with important consequences for low-energy hadronic
physics. As is well known, chiral symmetry breaking can be
understood through the celebrated Banks-Casher relation~\cite{BC} in
terms of the accumulation of small eigenvalues of the Dirac
operator. Similarly, the restoration of chiral symmetry at higher 
temperatures, in the deconfined phase of QCD, is related to the
depletion of this spectral region. 
The finite temperature transition in QCD is actually a
crossover~\cite{Aoki:2005vt,Borsanyi:2010cj}, with both the chiral and
confining properties of the theory undergoing a rapid change 
in a small interval near the pseudocritical temperature $T_c$.  
In recent years, it has become more and more clear that the decrease in
the spectral density of low modes is accompanied by a change in the
localisation properties of the corresponding eigenvectors: while at low
temperatures the low modes are extended throughout the whole spatial
volume, above the crossover temperature the lowest modes are
spatially localised on the scale of the inverse temperature. Evidence
for this behaviour has been obtained by means of lattice simulations
with different fermion discretisations, namely
staggered~\cite{GGO2,KP,BKS,KP2,feri,crit}, overlap~\cite{KGT,BKS}  
and domain wall~\cite{Cossu:2016scb} fermions.

While the existence of a close relation between deconfinement, chiral
symmetry restoration and localisation of the lowest modes has by now
been convincingly demonstrated, the physical mechanisms behind these 
phenomena and their interplay are still under study. An interesting
possibility is that one of the three phenomena is actually triggering
the other two: this 
would somehow reduce the need to find an explanation to the
``fundamental'' phenomenon only. There are several indications that
deconfinement might be such a ``fundamental'' phenomenon. 
Perhaps the clearest hint is that also in pure-gauge theories one
finds chiral symmetry restoration\footnote{Here chiral symmetry
  restoration (resp.~breaking) is understood as the spectral density 
  of the Dirac operator being zero (resp.~finite) at the origin.} and
localisation of the lowest Dirac eigenmodes at high temperature, above
the deconfinement transition. 
The first evidence that deconfinement, localisation and the chiral
transition happen around the same temperature in quenched QCD
came from Ref.~\cite{GGO2}. 
Moreover, the mechanism for localisation
proposed in Refs.~\cite{BKS,GKP,GKP2} is based on the presence of
``islands'', where the Polyakov lines fluctuate away from the ordered
(trivial) value, in the ``sea'' of ordered Polyakov lines. Such
``islands'' provide an ``energetically'' favourable location for the
eigenmodes. In a sense, localisation is thus ``reduced'' to
deconfinement. Support to this explanation was given in
Ref.~\cite{BKS} by studying the correlation of the Dirac
eigenfunctions with the fluctuations of the Polyakov loop on SU$(2)$
gauge configurations, and in Refs.~\cite{GKP,GKP2} by designing toy
models that should feature localisation precisely through the proposed
mechanism. Further evidence has been recently obtained in
Ref.~\cite{Cossu:2016scb}, in which a clear correlation is reported
between the position of localised modes and that of the favourable
``islands'' of Polyakov loops. 

Relating deconfinement and chiral symmetry restoration is more
difficult, despite the ample numerical evidence of a close
relationship. 
The connection between deconfinement on one side, and both
localisation and chiral symmetry restoration on the other has been
investigated in Ref.~\cite{GKP2} in the context of a QCD-inspired toy
model. This model is essentially obtained by replacing the Polyakov
lines with spin-like variables and by simplifying the dynamics of the
spatial gauge links. Only a few dynamical
properties of QCD are retained, namely the existence of an ordered
phase for the spins/Polyakov lines, with local disorder 
corresponding to ``islands'' of unordered spins/Polyakov lines; and the
correlation of spatial gauge links across time slices. 
Despite the drastic simplification, this toy model is able to
correctly reproduce the qualitative features of localisation and of
the chiral transition.
A detailed mechanism relating deconfinement and chiral restoration has
not been proposed in Ref.~\cite{GKP2}. Nevertheless, it is suggested
there that the depletion of the spectral region around the origin
depends on the presence of order in the Polyakov line configuration in
two ways. The accumulation of small eigenmodes requires both the
presence of sizeable fluctuations of the Polyakov line away from the
trivial value, and the possibility for the eigenmodes to effectively
mix different temporal-momentum components of the wave functions
throughout the spatial volume, which is hampered by the localisation
of the aforementioned fluctuations within disconnected ``islands''.

An alternative, or possibly complementary view of the relation between
the chiral transition and localisation of the lowest modes is obtained
from the picture of the QCD vacuum as an ensemble of topological
objects. The idea of a relation between localised modes and
topological objects was first put forward by Garc\'ia-Garc\'ia and 
Osborn in Ref.~\cite{GGO}, in the framework of the 
Instanton Liquid
Model~\cite{Shuryak:1981ff,Shuryak:1982dp,Shuryak:1982hk},
starting from the analogy between spontaneous chiral symmetry breaking
and conductivity in a disordered
medium~\cite{Diakonov:1984vw,Diakonov:1985eg,Diakonov:1995ea,Smilga:1992yp,
Janik:1998ki,Osborn:1998nm,Osborn:1998nf,GarciaGarcia:2003mn}. 
The basic idea is that the low modes of  
the Dirac operator are essentially coming out of the mixing of the
zero modes supported by instantons and anti-instantons (or, more
precisely, by their finite-temperature analogues, namely the
calorons). At finite temperature, the matrix elements of the Dirac
operator between these zero modes decay exponentially with the
distance between the topological objects, so that one is effectively
dealing with a random system with short-distance interactions. The
density of topological objects plays the role of the amount of
disorder in the system, and at a certain critical value of the density
an Anderson transition takes place, with localisation of the lowest
modes, and the opening of a gap in the spectral density around the
origin. Even though the Instanton Liquid Model does not seem
to provide an adequate description of deconfinement at the QCD
transition, nevertheless the mechanism proposed in
Ref.~\cite{GGO} could be at work, with
instantons/anti-instantons replaced by the 
appropriate, zero-mode supporting topological objects. In this
respect it is worth mentioning the results of
Ref.~\cite{Cossu:2016scb}, which support a connection between
localised modes and certain topological objects, namely
the monopole-instantons, which might be
responsible for the deconfinement transition (see 
Refs.~\cite{Diakonov:2009jq,Bruckmann:2003yq,Shuryak:2012aa,Poppitz:2012nz} 
and references therein). This line of studies
certainly deserves further attention. 

Even if the mechanisms mentioned above are correct, the fact that the
QCD deconfinement/chiral transition is actually an analytic crossover
implies that they act somehow gradually, and so telling which
phenomenon is the ``fundamental'' one becomes a not so well defined
question. For this reason, it is interesting to study QCD-like
models with a more clear-cut situation, namely models displaying a
genuine phase transition, where one could check if the three phenomena
take place together, and possibly tell what is triggering what. This
would certainly help in understanding the interplay of deconfinement,
chiral restoration and localisation in the physical case of QCD.

A study of this kind appeared in Ref.~\cite{GGO2}. Their findings
indicate that deconfinement, chiral symmetry restoration and onset of
localisation near the origin all take place around the same
temperature in the pure-gauge SU$(3)$ theory. 
Although it seems unlikely that the three phenomena occur at nearby
but different temperatures, especially given the presence of a
temperature where confining and chiral properties change abruptly, in
order to make a stronger statement one should improve on the not so
large lattice sizes employed in that paper.

Another useful model for the study of these issues is provided by
unimproved staggered fermions on lattices with temporal extension
$N_T=4$ (in lattice
units)~\cite{unimproved0,unimproved1,unimproved}. Differently from 
pure-gauge SU$(3)$, this model possesses a true (albeit softly broken)
chiral symmetry. In this respect this model is closer to QCD, and
chiral symmetry restoration is here a better defined issue.
Moreover, as a statistical physics system, this model displays a genuine,
first-order phase transition, where the confining and chiral
properties~\cite{unimproved0,unimproved1,unimproved}, and as we will
see also the localisation properties, all undergo a sharp change. 
The study of this model is the subject of this paper, which is
organised as follows. In Section \ref{sec:loc} we briefly review
the phenomenon of localisation in high-temperature QCD, and we discuss
in particular how one can conveniently detect it. In Section
\ref{sec:num}, after a brief description of the model under scrutiny,
we provide numerical evidence that in this model deconfinement, chiral
symmetry restoration and localisation of the lowest modes take place 
simultaneously. Preliminary results have already been reported 
in Ref.~\cite{GKKP}. Finally, in Section \ref{sec:concl} we discuss
our conclusions and prospects for the future.  

\section{Localisation in high-temperature QCD}
\label{sec:loc}

In this section we review localisation in lattice QCD 
at high temperature, and discuss in general how one can detect
it. More details can be found in the original references and in the
review paper Ref.~\cite{Giordano:2014qna}.   

While at low temperatures the low-lying eigenmodes of the Dirac
operator are delocalised on the entire lattice
volume~\cite{VWrev,deF}, above the crossover temperature,
$T_c$~\cite{Aoki:2005vt,Borsanyi:2010cj}, they become spatially
localised~\cite{GGO2,KGT,KP,BKS,KP2,feri,crit,Cossu:2016scb} 
on the scale of the inverse temperature~\cite{KP,KP2,Cossu:2016scb}.
From now on we focus on staggered fermions, both for simplicity and
because of the larger amount of available evidence. Since in this case
the eigenvalues $i\lambda$ are purely imaginary and the spectrum is
symmetric with respect to zero, it suffices to discuss $\lambda \ge
0$. For results about overlap fermions compare Refs.~\cite{KGT,BKS},
while recent results concerning domain-wall fermions can be found in
Refs.~\cite{Cossu:2016scb,Tomiya:2016jwr}. 

In high temperature QCD, eigenmodes corresponding to eigenvalues below a
temperature-dependent critical point in the spectrum, the ``mobility
edge'' $\lambda_c=\lambda_c(T)$, are localised in a finite region of the
lattice. Eigenmodes above $\lambda_c$, on the other hand, occupy the
whole lattice volume. The curve $\lambda_c(T)$ reaches zero 
at a temperature compatible with $T_c$~\cite{KP2}, as determined from
thermodynamic observables~\cite{Aoki:2005vt,Borsanyi:2010cj}. The
transition in the spectrum from localised to delocalised modes, taking
place at the critical point $\lambda_c$, was shown to be a genuine
second-order phase transition~\cite{crit}, analogous to the
metal-insulator transition in the Anderson
model~\cite{Anderson58,LR,EM}, which describes non-interacting
electrons in a disordered crystal. Furthermore, in Ref.~\cite{crit}
the correlation-length critical exponent was found to be compatible
with that of the 3D unitary Anderson model~\cite{nu_unitary}. Here
``unitary'' refers to the symmetry class of the model in the Random
Matrix Theory (RMT)
classification of random matrix ensembles~\cite{Mehta}, which is
shared by the staggered Dirac operator~\cite{VWrev}. A study of the
multifractal properties of the eigenmodes at criticality~\cite{UGPKV}
confirmed the result for the critical exponent, and also showed
that the multifractal exponents of the critical eigenmodes in QCD are
compatible with those of the 3D unitary Anderson model~\cite{UV}.
This provided further evidence that the delocalisation
transitions in the two models belong to the same universality class. 

We finally mention that localisation of the lowest eigenmodes has been
observed also in QCD-like theories, like SU$(2)$ pure-gauge theory with
staggered~\cite{BKS,KP} or overlap fermions~\cite{BKS}, and SU$(3)$
pure-gauge theory~\cite{GGO2}. In all these cases, fermions in the
fundamental representation were considered. In these models the
deconfinement/chiral transition is a genuine phase transition, and 
localisation of the lowest modes is present only in the
high-temperature phase. In the case of SU$(3)$ pure-gauge theory there
is also evidence that localisation appears near the critical
temperature. 

We now want to discuss how one can determine the presence of localised
modes in the Dirac spectrum. 
The most direct way of detecting localisation is of course to study
the amount of spatial volume occupied by a given eigenmode, and how
this scales with the lattice size. A convenient observable is the
so-called participation ratio ($PR$), defined for a given normalised 
eigenmode $\psi_n$ as
\begin{equation}
  \label{eq:pr_def}
  PR_n = \f{1}{N_T V}(IPR_n)^{-1} = \f{1}{N_T V} \left[\sum_{\vec x,t}|
    \psi_n^\dag(t,\vec x) 
  \psi_n(t,\vec x)|^2\right]^{-1}\,,
\end{equation}
where $IPR$ stands for ``inverse participation ratio'', and
$\psi_n^\dag \psi_n =  \sum_a (\psi_n)_a^* (\psi_n)_a$ stands
for summation over the colour degree of freedom. Here $N_T$ is the
temporal extension of the lattice and $V=L^3$ the spatial volume.
In the infinite-volume limit, the $PR$ tends to some finite constant
for delocalised modes, while it goes to zero for localised modes.

A convenient shortcut to the average localisation properties of the
eigenmodes in a given spectral region is provided by the statistical
properties of the corresponding eigenvalues. Indeed, delocalised modes
are expected to be freely mixed by fluctuations of the gauge fields,
and so the corresponding eigenvalues are expected to obey the
statistics of the appropriate ensemble of RMT,
which for staggered fermions is the Gaussian Unitary
Ensemble~\cite{VWrev}. Localised modes, on the other hand, are
sensitive only to fluctuations of the gauge fields taking place where
they are located, and so the corresponding eigenvalues are expected to
fluctuate independently, thus obeying Poisson statistics. In the case 
of both RMT and Poisson statistics, analytic predictions are available
for the so-called unfolded spectrum~\cite{Mehta}, obtained by means of
a local rescaling of the eigenvalues which leads to unit spectral
density uniformly through the spectrum. In practice, unfolding is
performed by sorting all the 
eigenvalues obtained on the available configurations, and replacing
them by their rank divided by the number of configurations.
In particular, the probability distribution $P_{\lambda}(s)$
of the unfolded level spacings $s_j = \f{\lambda_{j+1}-\lambda_j}{\la
  \lambda_{j+1}-\lambda_j\ra_\lambda}$ is known for both kinds of
statistics. Here $\la \lambda_{j+1}-\lambda_j\ra_\lambda$ is the average level
spacing in the spectral region corresponding to the level $\lambda_j$,
and the subscript $\lambda$ means that $\la
\lambda_{j+1}-\lambda_j\ra_\lambda$ and $P_{\lambda}(s)$ are computed
locally in the spectrum. In general, this is done by dividing the
spectrum in disjoint bins of fixed size $w$, averaging the desired
observable within a bin, and assigning the result to the average
$\lambda$ in that bin. 

The transition from localised to delocalised
modes can then be detected by measuring the statistical properties of
the unfolded spectrum locally and comparing them to the analytical
results. A particularly convenient observable in this respect is the
integrated probability distribution function,
$I_{s_0}(\lambda)$~\cite{SSSLS,HS},   
\begin{equation}
  \label{eq:i05_def}
  I_{s_0}(\lambda) \equiv \int_0^{s_0} ds\,P_{\lambda}(s)\,,
\end{equation}
which has clearly distinguishable values for Poisson and RMT
statistics if $s_0$ is chosen properly. The difference between the
Poisson and the RMT values is maximised by $s_0\simeq 0.508$, which is
the point closest to $s=0$ where the two distributions cross. In this
case $I_{s_0}^{\rm Poisson}\simeq 0.398$ and $I_{s_0}^{\rm RMT}\simeq
0.117$. In a finite volume $I_{s_0}(\lambda)$ interpolates smoothly
between the two limiting values, with the transition becoming sharper
as the volume is increased. At the ``mobility edge'', $\lambda_c$,
$I_{s_0}(\lambda)$ is volume independent~\cite{SSSLS,HS}, and takes
the critical value $I_{s_0}^{\rm crit}$. In Ref.~\cite{crit} this was 
determined to be $I_{s_0}^{\rm crit}~\simeq 0.1966(25)$ in QCD with
$2+1$ flavours of staggered fermions. Since $I_{s_0}^{\rm crit}$ is
determined by the critical eigenvalue statistics, which is believed to
be universal as well as volume independent~\cite{SSSLS}, this value
should be attained at $\lambda_c$ in all those models displaying a
localisation transition along the spectrum which belongs to the same
universality class as that of the 3D unitary Anderson model. Since
only the dimensionality of the model and the symmetry class matter for
the universality class, we expect $I_{s_0}^{\rm crit}$ to always be
the critical value of $I_{s_0}$ in QCD-like models at finite
temperature, as long as we use staggered fermions and gauge group
SU$(N_c\ge 3)$. For a detailed discussion of why these models have to
be considered three-dimensional in this context, the reader can confer
Refs.~\cite{GKP,GKP2}. The result obtained in QCD can then be used
in all these models to determine $\lambda_c$ as the point in the 
spectrum where $I_{s_0}(\lambda)$ takes the critical value. This can
be done by using configurations corresponding to a single lattice
volume. This approach is simpler than using the strict definition of
$\lambda_c$ as the point in the spectrum where $I_{s_0}(\lambda)$ is
volume-independent, but not as precise, although it will eventually be
the same in the thermodynamic limit. It is worth 
noting that any definition of $\bar\lambda_c$ as the point in the
spectrum where $I_{s_0}(\bar\lambda_c)=\bar{I}_{s_0}$ with 
$ I_{s_0}^{\rm RMT}< I_{s_0}< \bar{I}_{s_0}^{\rm Poisson} $ will eventually
converge to the right value in the thermodynamic limit, but with
different finite size effects. Choosing the critical value is likely
to minimise these effects.

Although the study of the statistical properties of the unfolded
spectrum is convenient for the detection of the
localisation/delocalisation transition in the spectrum, it is not
quite appropriate to determine $\lambda_c$ when this is very close to
the origin. This is a consequence of the fact that in QCD and similar
theories the spectral density is small and rapidly varying in the
localised part of the spectrum. In turn, this makes unfolding
unreliable in that spectral region in a finite volume. Since
$\lambda_c$ tends to zero as one gets close to the critical
temperature, its determination as described above is thus affected by
uncontrolled systematic errors near the critical temperature. Using
the volume-independence of the statistical properties of the spectrum
would not improve the situation either.  

In order to assess whether the chiral transition and the appearance of
localised modes take place at the same temperature, it is more
convenient to employ other observables. Possibly the simplest way to
detect the onset of localisation in the low end of the spectrum is the
study of the average participation ratio of the lowest eigenmode,
corresponding to the lowest eigenvalue, $\lambda_1$. This was the
strategy employed in Ref.~\cite{GGO2}. If localisation and chiral
restoration happen together, we expect the participation ratio of the
smallest mode to change from being of order 1 to some small value,
which tends to zero as the lattice size is increased. 

The distribution of the lowest eigenmode provides also a way to
cross-check the presence of a chiral transition, or more precisely of
a jump in the spectral density at the origin at some critical
temperature. If chiral symmetry is broken, the
spectral density is expected to be finite near the origin, and the
small eigenmodes are expected to obey chiral random matrix theory
(chRMT)~\cite{VWrev}. More precisely, chRMT makes definite
predictions for the statistical properties of the microscopic
spectrum, $z$, defined by rescaling the eigenvalues with the spectral
density at the origin, $z = \lambda \pi \rho(0)$. In particular, the
distribution of the smallest eigenvalue in a given topological sector
is known analytically: for example, the smallest rescaled eigenvalue
$z_1 = \lambda_1 \pi \rho(0)$ in the trivial topological sector and in
the quenched theory is expected to be distributed according to the
following probability distribution
function~\cite{Forrester,Nishigaki:1998is,Nishigaki:2016nka}: 
\begin{equation}
  \label{eq:rmt_nu0}
  P_1^{{\rm chRMT}, {\rm quenched}, \nu=0}(z_1) = \f{z_1}{2} e^{-\left(\f{z_1}{2}\right)^2}\,.
\end{equation}
The important point is that since $\rho(0)$ is
proportional to the lattice volume, $V$, then the $n$-th moment of the
distribution of the smallest eigenvalue will scale like $V^{-n}$. For
example, $\la z_1 \ra_{{\rm quenched}, \nu = 0} = \sqrt{\pi}$, which leads to $\la
\lambda_1 \ra_{\nu=0} = [\sqrt{\pi}\rho(0)]^{-1}$; in general, one has $\la
\lambda_1 \ra \sim V^{-1}$. 
On the other hand, at high temperature the eigenmode corresponding to
the smallest eigenvalue is expected to be localised, and thus 
the distribution of the smallest eigenvalue is determined by the spectral
density and the Poisson statistics of the localised eigenmodes. 
Assuming a power-law behaviour
$\rho(\lambda)= CV\lambda^\alpha$ for the spectral density near the
origin, one can write down explicitly the distribution of the smallest
eigenvalue~\cite{KGT}:
\begin{equation}
  \label{eq:smallest_ev_hT}
  P_1(\lambda) = CV\lambda^{\alpha} e^{-\f{CV}{\alpha+1}\lambda^{\alpha+1}}\,.
\end{equation}
Such a power-law behaviour has been indeed observed for staggered
fermions~\cite{KP2}. From this it immediately follows that~\cite{KGT}
\begin{equation}
  \label{eq:smallest_ev_hT_2}
  \la\lambda_1\ra =
  \left(\f{CV}{1+\alpha}\right)^{-\f{1}{1+\alpha}}
  \Gamma\left(1+\f{1}{1+\alpha}\right) \sim V^{-\f{1}{1+\alpha}}\,.
\end{equation}
In conclusion, if chiral symmetry restoration and the onset of
localisation take place together at some critical $\beta_c$, one
expects that at that point also the scaling with the volume of $\la
\lambda_1\ra$ will change. There are two technical details which are
worth mentioning. In a finite volume one necessarily has vanishing
spectral density at the origin, so in rescaling the eigenvalues to
obtain the microscopic spectrum at low temperature one has to use the
infinite-volume limit of $\rho(0)/V$. 
At high temperature, Poisson
statistics for the lowest modes in the case of staggered fermions is
distorted by the presence of ``doublets'', i.e., pairs of close
eigenvalues. Although this could change
Eq.~\eqref{eq:smallest_ev_hT_2}, we still expect that
$\la\lambda_1\ra$ vanishes with the volume faster than $1/V$.

\begin{figure}[t]
  \centering
  \includegraphics[width=0.8\textwidth]{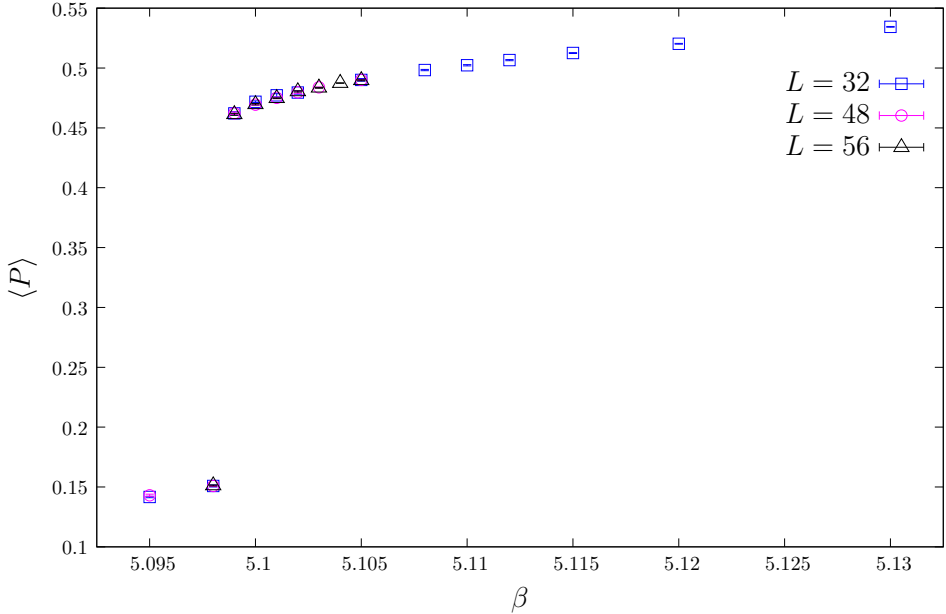}
  \caption{Average Polyakov loop as a function of $\beta$.}
  \label{fig:poly}
\end{figure}

\section{$N_T=4$ unimproved staggered fermions}  
\label{sec:num}

In this section, after briefly describing the model of interest, we
provide several pieces of evidence that unimproved staggered fermions
on lattices with temporal extension $N_T=4$ display localisation in
the deconfined/chirally restored phase, and that the appearance of
localised modes takes place simultaneously with the corresponding
first-order phase transition.

\subsection{The model}
\label{sec:model}

The model we have studied consists of $N_f=3$ degenerate lattice
fermions in the fundamental representation, interacting via SU$(3)$
gauge fields. The staggered discretisation without improvement is used
for the fermions, together with the rooting trick, and the Wilson
action is used for the gauge fields. The temporal extension of the
lattice is fixed to $N_T=4$. The partition function thus reads
\begin{equation}
  \label{eq:part_func}
  Z = \int DU \left\{\det \left[D_{\rm stag}(U)+
      m\right]\right\}^{\f{3}{4}} e^{-\beta S_{\rm W}(U)}\,, 
\end{equation}
where $D_{\rm stag}$ is the staggered Dirac operator, $S_{\rm W}$ the
Wilson action, $U$ are SU$(3)$ matrices living on the lattice links,
and $DU$ denotes the product of the corresponding Haar measures. 
Moreover, $m$ is the (bare) fermion mass and $\beta$ the gauge
coupling. 

The relevant symmetries of this model are an SU$(3)$ chiral symmetry,
softly broken by the fermion mass term, and a $\mathbb{Z}_3$ center
symmetry, broken by the presence of fermions. It is
known~\cite{unimproved0,unimproved1,unimproved} that this model
displays a first-order deconfining and chirally-restoring phase
transition, for bare quark masses below the critical value
$m<0.0259$~\cite{unimproved}. More precisely, what one observes is a
finite jump in the relevant order parameters, namely the Polyakov loop
and the chiral condensate, as the gauge coupling is increased beyond a
critical value. From the point of view of QCD the presence of a
genuine phase transition rather than a crossover is just a lattice
artifact, caused by the coarseness of the lattice, and it does not
survive the continuum limit. Nevertheless, one can treat this system
as a statistical mechanics model, and study how its properties change
when changing the gauge coupling. 

\begin{figure}[t]
  \centering
  \includegraphics[width=0.8\textwidth]{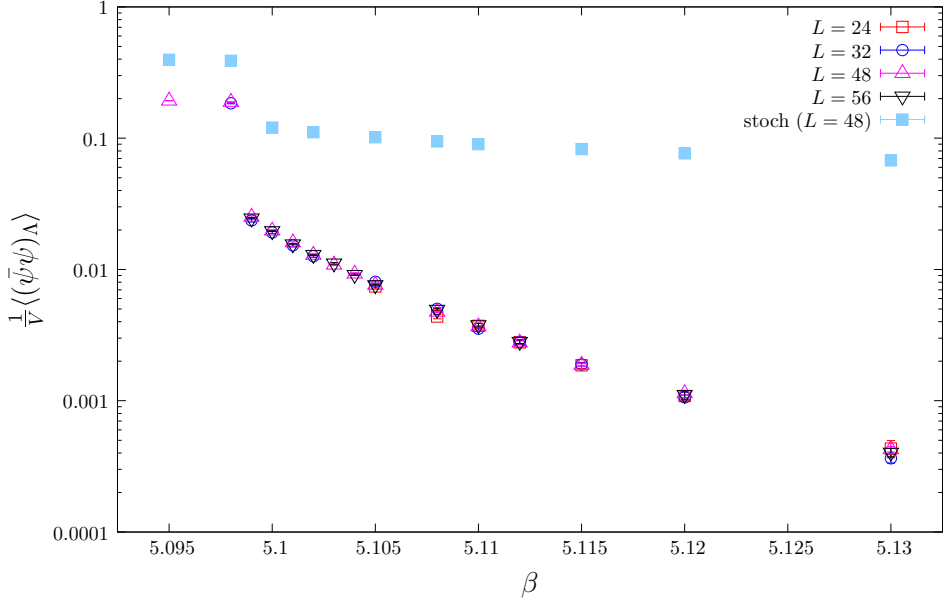}
  \caption{Chiral condensate, as defined in
    Eq.~\protect{\eqref{eq:chico}}, divided by the spatial volume, as
    a function of $\beta$ (logarithmic scale). The full chiral
    condensate for $L=48$ is also shown for comparison.}  
  \label{fig:cond}
\end{figure}

\subsection{Numerical results}
\label{sec:numres}

In this section we present our numerical results. Simulations have
been carried out on lattices of spatial size $L=24,32,48,56$, in a range
of couplings $\beta = 5.095 \div 5.130$. The bare fermion mass is set to
$m=0.01$, well below the critical value. Here and in the following all
quantities are in lattice units. 

We begin by showing that a deconfining and chiral-symmetry-restoring
first-order phase transition takes place at some critical
$\beta_c$. 
In Fig.~\ref{fig:poly} we show the expectation value of the Polyakov
loop $\la P \ra$, where $P(\vec x)=\f{1}{3}\Re \tr \tilde P(\vec x)$,
and $\tilde P(\vec x)$ is the Polyakov line, i.e., the straight-line
gauge transporter winding around the temporal direction at $\vec x$. A
sudden jump is clearly 
visible between $\beta=5.098$ and $\beta=5.099$, indicating the
presence of a first-order phase transition. The critical coupling
for deconfinement, $\beta_c^{\rm dec}$, is therefore in the
window $5.098<\beta_c^{\rm dec}<5.099$.

Next, in Fig.~\ref{fig:cond} we show the chiral condensate
$\la(\bar\psi\psi)_\Lambda\ra$, defined as
\begin{equation}
  \label{eq:chico}
  \la(\bar\psi\psi)_\Lambda\ra = \int_0^\Lambda d\lambda\,
  \f{2m}{\lambda^2 + m^2}\,\rho(\lambda)\,.
\end{equation}
We used $\Lambda=0.001$. It is evident that between
$\beta=5.098$ and $\beta=5.099$ also this quantity jumps abruptly,
decreasing by an order of magnitude. The critical coupling where the
chiral transition takes place, $\beta_c^\chi$, is therefore 
in the same window $5.098<\beta_c^\chi<5.099$ as $\beta_c^{\rm dec}$. 
It is then very likely that $\beta_c^\chi=\beta_c^{\rm dec}$.
Although the value we used for $\Lambda$ seems unreasonable as it is
much smaller than the fermion mass, the effect that we are after,
namely the presence of a first-order phase transition, should not
depend on our choice for the cutoff. In fact, what we expect to cause
the effect is a change in $\rho(0)$, i.e., in the density of the
lowest modes. In Fig.~\ref{fig:cond} we also show
the full chiral condensate (i.e., $\Lambda=\infty$) computed
stochastically for one of the volumes. While the two quantities are
numerically quite different, they nevertheless show the same
critical behaviour, and so $\la(\bar\psi\psi)_\Lambda\ra$ can be
reliably used to infer the presence of a phase transition.

\begin{figure}[t]
  \centering
  \includegraphics[width=0.8\textwidth]{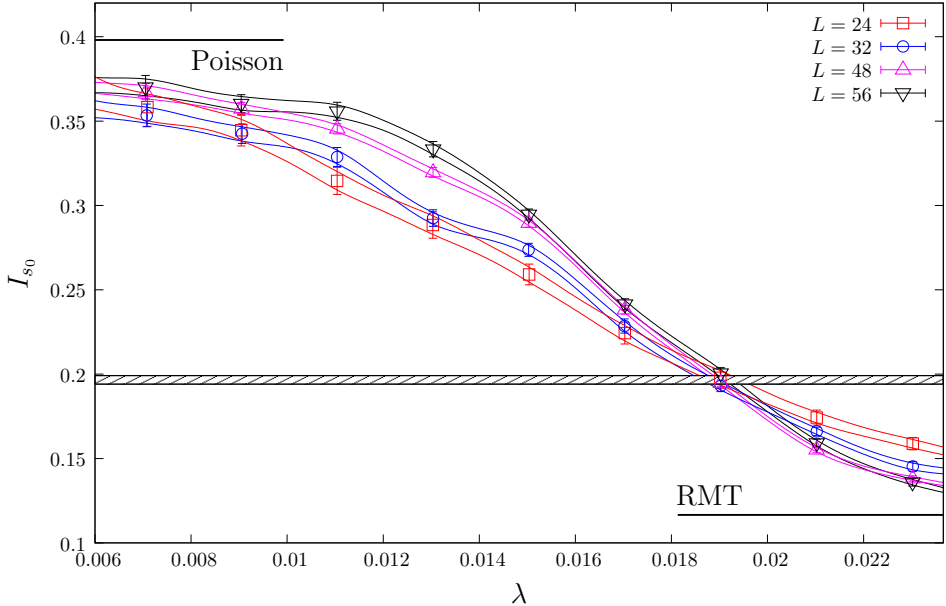}
  \caption{Local spectral statistics $I_{s_0}$ as a function of
    $\lambda$ in the deconfined phase at $\beta=5.130$. Both numerical
    data and spline 
    interpolations are shown. The values 
    corresponding to Poisson and RMT are also shown. The value
    corresponding to the critical statistics, as determined in
    Ref.~\cite{crit}, is also shown together with its error
    band. }
  \label{fig:0}
\end{figure}

We now discuss the localisation properties of the eigenmodes. First of
all, we show in Fig.~\ref{fig:0} the typical behaviour of the 
spectral statistic $I_{s_0}$ as one moves along the spectrum in the
deconfined phase. Here $\beta$ is fixed at $\beta=5.130$, and we scan
the spectrum. Moving from the low end of the spectrum towards the
bulk, a clear transition from Poisson to RMT statistics is
observed. This shows that the lowest modes are localised, while higher
up in the spectrum the eigenmodes are extended.
Moreover, $I_{s_0}$ passes through the critical value where
it is volume-independent (within errors), as it should be. This
supports the expected universality of the critical statistics.

\begin{figure}[t]
  \centering
  \includegraphics[width=0.8\textwidth]{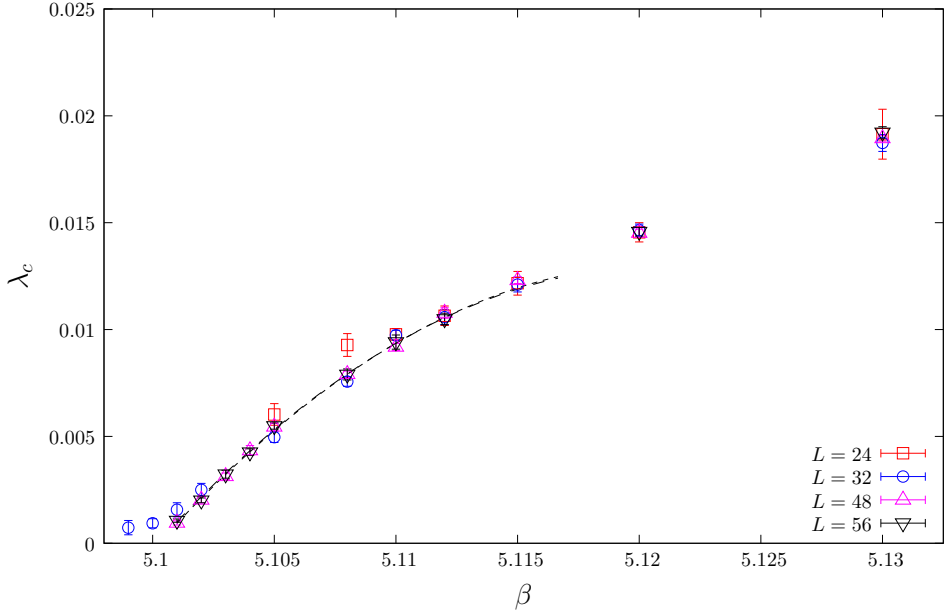}
  \caption{Mobility edge $\lambda_c$, determined as the point where
    $I_{s_0}(\lambda_c)=I_{s_0}^{\rm crit}$, as a function of
    $\beta$. Quadratic fits to the data are also reported, for $L=48$
    (long-dashed line) and $L=56$ (short-dashed line), with almost
    identical curves.} 
  \label{fig:3}
\end{figure}

The volume-independence of $I_{s_0}$ where it crosses the critical
value, $I_{s_0}^{\rm crit}$, shows that the latter can be used to
reliably determine $\lambda_c$ via the relation 
$I_{s_0}(\lambda_c)=I_{s_0}^{\rm crit}$, as discussed in section
\ref{sec:loc}. Our results for $\lambda_c(\beta)$ determined in this
way are shown in Fig.~\ref{fig:3}. 
Here the bin size for $\lambda$ is $w=0.001$.
To find $\lambda_c$ we used a
cubic spline interpolation $f(\lambda)$ of the numerical data. The
error on the spline was determined as follows. For fixed
$\{\lambda_i\}$, we generated 1000 synthetic sets of data points $\{I^{\rm
  synth}_{s_0}(\lambda_i)\}$, following a Gaussian distribution 
centered at $I_{s_0}(\lambda_i)$ and with standard deviation given by
the corresponding error, and then computed the corresponding set of
interpolating functions. The error $\delta f(\lambda)$ on the spline
interpolation at a given $\lambda$ was then computed as the standard
deviation of the set of such interpolating functions. 
We then determined the crossing points $\lambda_\pm$ by solving
$f(\lambda_\pm) \pm \delta f(\lambda_\pm) =
I_{s_0}^{\rm crit}$. The value of $\lambda_c$ was then
determined as the average of $\lambda_\pm$, and the corresponding
difference constitutes a first contribution to the uncertainty on
$\lambda_c$. A second source of error is the uncertainty on the value
of $I_{s_0}^{\rm crit}$, which has been determined numerically in
Ref.~\cite{crit}. To take this into account, we 
repeated the determination of $\lambda_\pm$ using 1000 synthetic
values of the critical $I_{s_0}$, generated according to a Gaussian
distribution with mean equal to $I_{s_0}^{\rm crit}$ and standard
deviation equal to the corresponding error. The final error on
$\lambda_c$ was obtained by adding in quadrature the two sources of
error. In Fig.~\ref{fig:3} we also show fits to the data (with $L=48$ 
or $L=56$, and up to $\beta=5.112$) with a second-order polynomial,
which yields for the critical point $\beta_c^{\rm loc}$ where
localisation appears the value $\beta_c^{\rm loc}\simeq 5.1$. This
value is quite close to $\beta_c^\chi$ and $\beta_c^{\rm
  dec}$. However, we cannot give a reliable estimate of the error
on $\beta_c^{\rm loc}$, since it is difficult to control the
systematic effects due to finite size on the determination of
$\lambda_c$ in the vicinity of the transition, when this is very close
to zero (see the discussion in section \ref{sec:loc}). 
Therefore, the method employed here does not allow us to determine
$\beta_c^{\rm loc}$ with the same accuracy with which $\beta_c^\chi$ and
$\beta_c^{\rm dec}$ can be obtained from the chiral condensate and the
Polyakov loop, respectively.

 \begin{figure}[t]
   \centering
   \includegraphics[width=0.8\textwidth]{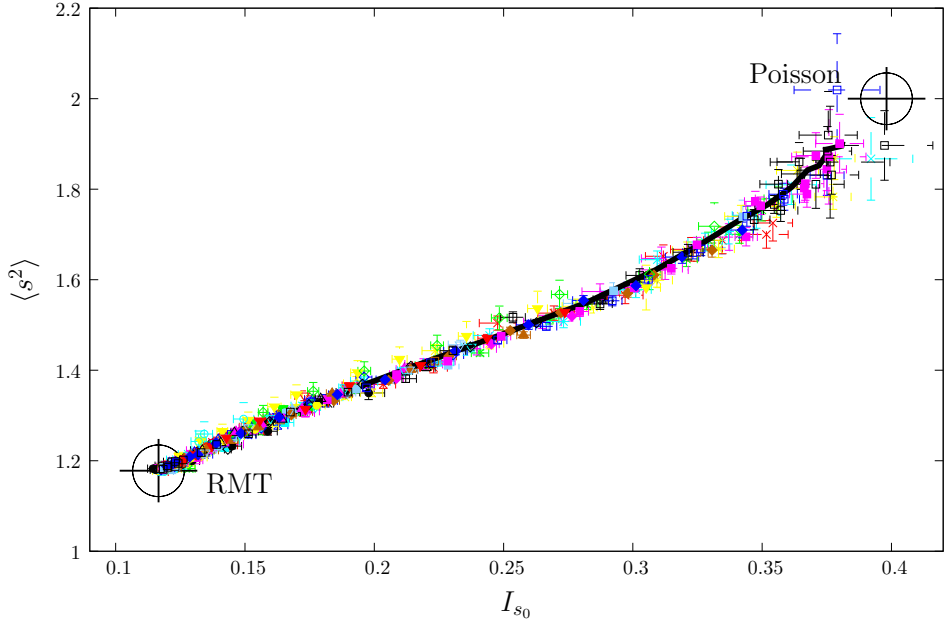}
   \caption{Shape analysis for $N_T=4$ unimproved staggered fermions
     for several volumes and values of the gauge coupling (points),
     and for $N_f=2+1$ QCD at $\beta=3.75$, $N_T=4$ and $L=56$ (solid
     line; data taken from Ref.~\cite{crit}). The points
     corresponding to Poisson and RMT statistics are also shown.}
   \label{fig:shape}
 \end{figure}

Before discussing other methods to determine $\beta_c^{\rm loc}$, it
is worth comparing our results for the statistical properties of 
the spectrum with those obtained in QCD, in order to test whether
there is a somewhat wider universality than just at the critical point
$\lambda_c$ in the spectrum. As discussed in
Ref.~\cite{Giordano:2014qna}, in QCD the unfolded level spacing
distributions $P_{\lambda}(s)$ found in different parts of the
spectrum lie on a universal path in the space  
of probability distributions, a path that is independent of volume,
temperature and lattice spacing. This can be seen by plotting two
different parameters of $P_{\lambda}(s)$ against each other (``shape
analysis''~\cite{shape_analysis}), 
thus taking a two-dimensional projection of this path, which should
therefore yield a universal curve. In Fig.~\ref{fig:shape} we plot the
second moment of the unfolded spacing distribution, $\la s^2\ra$,
against $I_{s_0}$, with each data point corresponding to a specific
point in the spectrum, and to a given system size and value of the
gauge coupling. The data points indeed arrange themselves rather
precisely on a single curve, which furthermore compares well with the
one obtained in QCD~\cite{Giordano:2014qna,Nishigaki:2013uya}. 

\begin{figure}[t]
  \centering
  \includegraphics[width=0.8\textwidth]{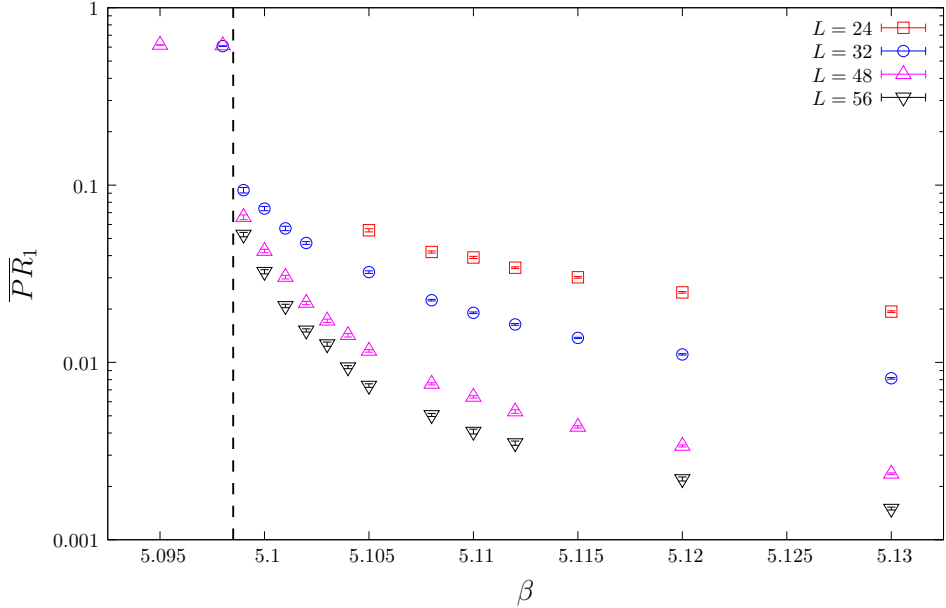}
  \caption{Average participation ratio  $\overline{PR}_1$
    of the first eigenmode as a function of $\beta$. The dashed line
    at $\beta=5.0985$ is halfway through the region where the
    deconfining/chiral transition is expected to take place.} 
  \label{fig:2}
\end{figure}

\begin{figure}[t]
  \centering
  \includegraphics[width=0.8\textwidth]{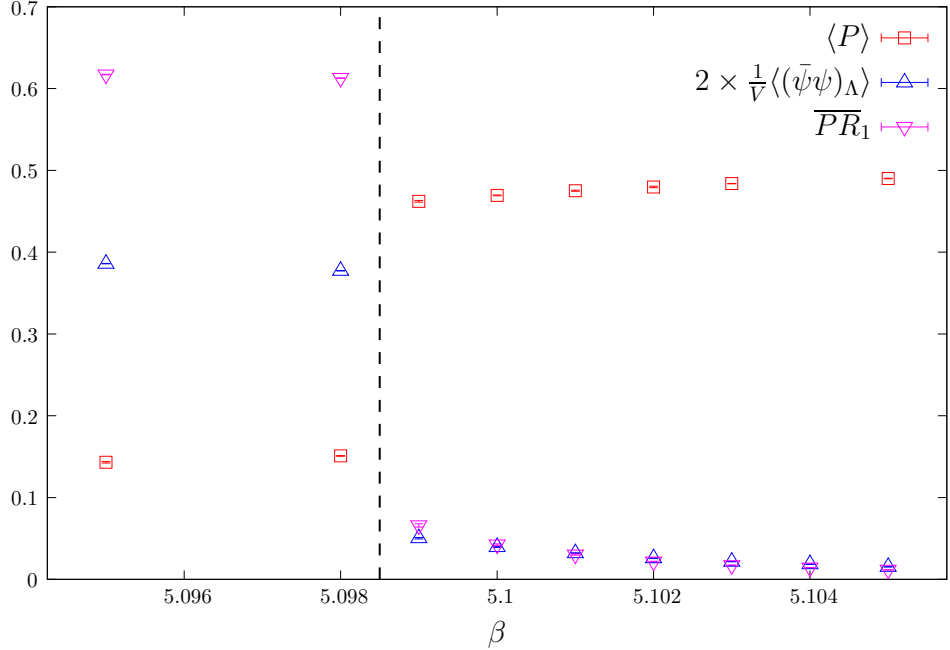}
  \caption{Average Polyakov loop $\la P\ra$, chiral condensate
    $\la(\bar\psi\psi)_\Lambda\ra $ and average participation ratio
    $\overline{PR}_1$ of the first eigenmode, as a function of
    $\beta$. The coincidence of the jump in the three observables is
    evident.}  
  \label{fig:2extra}
\end{figure}

As we mentioned above and in the previous Section, the determination of
$\lambda_c$ is difficult in the vicinity of the phase transition.
For this reason, we have studied in detail the behaviour of the first
eigenmode. In Fig.~\ref{fig:2} we show the participation ratio 
$\overline{PR}_1=(N_T V)^{-1}\la  (IPR_1)^{-1}\ra$  of the
first eigenmode averaged over configurations, as a function of
$\beta$. This plot shows clearly that from $\beta=5.099$ up the 
first eigenmode is localised, with $\overline{PR}_1$ tending to zero
as the volume is increased. This indicates that $5.098<\beta_c^{\rm
  loc}<5.099$, so that our results are compatible with 
$\beta_c^{\rm loc}=\beta_c^{\rm dec}=\beta_c^\chi$. For completeness,
we thus show the Polyakov loop, the chiral condensate and the
participation ratio of the first eigenmode together in
Fig.~\ref{fig:2extra}.

\begin{figure}[t]
  \centering
  \includegraphics[width=0.8\textwidth]{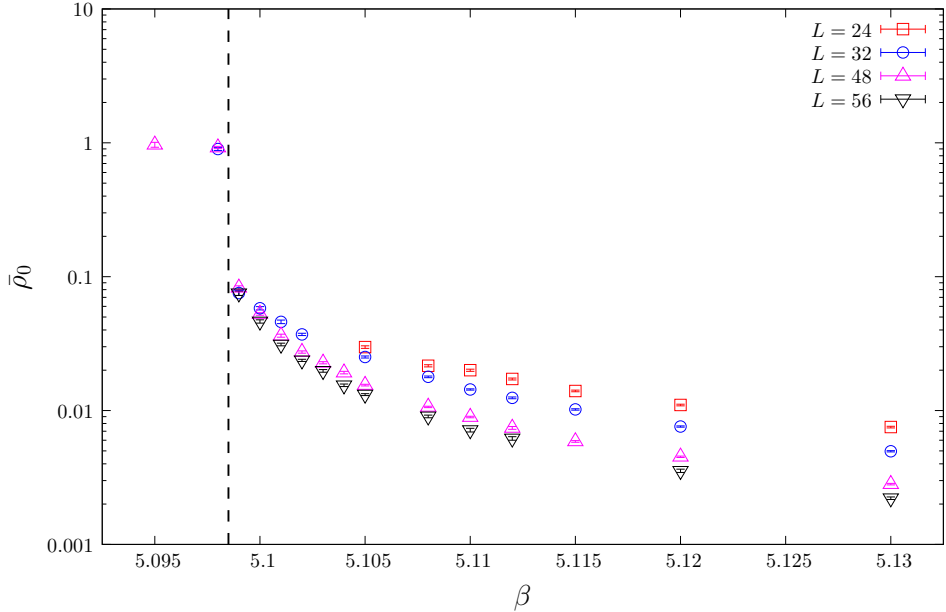}
  \caption{Plot of $\bar\rho_0$, Eq.~\protect{\eqref{eq:barrho_def}},
    as a function of $\beta$.}
  \label{fig:1}
\end{figure}

\begin{figure}[t]
  \centering
  \includegraphics[width=0.8\textwidth]{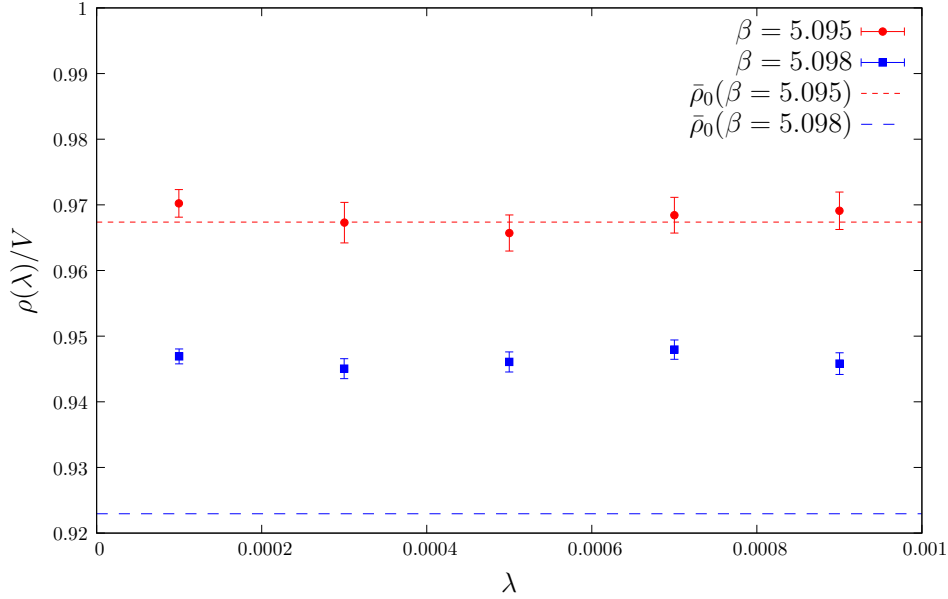}
  \caption{Comparison of the spectral density per unit volume with
    $\bar\rho_0=[V \sqrt{\pi}\la \lambda_1\ra ]^{-1}$ in the chirally
    broken phase. Here $L=48$.}
  \label{fig:barrho}
\end{figure}

As a cross-check for the coincidence of $\beta_c^{\chi}$ and
$\beta_c^{\rm loc}$, in Fig.~\ref{fig:1} we show the quantity
$\bar\rho_0$, defined as
\begin{equation}
  \label{eq:barrho_def}
\bar\rho_0\equiv \f{1}{V \sqrt{\pi}\la \lambda_1\ra}\,.  
\end{equation}
If the theory were quenched and only the topological sector $\nu=0$
contributed to the partition function, this would yield the spectral
density at the origin, in case this were not vanishing. If the
spectral density vanished at the origin like 
some power of $\lambda$, then this quantity would also vanish in the
infinite-volume limit [see Eq.~\eqref{eq:smallest_ev_hT_2}]. We
therefore expect this quantity to behave like $\rho(0)$ across the
transition, namely to tend to a finite constant below $T_c$, and to
zero above $T_c$, when the thermodynamic limit is taken. This
expectation holds if the eigenvalues obey RMT and Poisson statistics,
respectively, which in turn should be the case for delocalised and
localised modes, respectively. In Fig.~\ref{fig:1} one can clearly see
a jump between $\beta=5.098$ and $\beta=5.099$, where $\bar\rho_0$
changes by an order of magnitude. Moreover, one can see that above
$\beta=5.099$, $\bar\rho_0$ tends to zero as the volume
increases. Despite it being above the jump, at $\beta=5.099$ one finds
that $\bar\rho_0$ is constant within errors. 
Although the current statistical accuracy and
the limited number of available volumes do not allow a definitive
conclusion, this behaviour is most likely a finite-size
effect. Indeed, as long as the volume is not larger than the typical 
size of the localised modes, these effectively look delocalised.
In conclusion, even though we cannot assign this point to either phase with
certainty, nevertheless the presence of a jump strongly suggests that
it belongs to the localised phase. This conclusion would be
consistent with the behaviour of $\overline{PR}_1$, 
shown in Fig.~\ref{fig:2}. In Fig.~\ref{fig:barrho} we show that
$\bar\rho_0$ compares indeed quite well with $\rho(\lambda\approx 0)$
in the chirally broken phase. The use of the quenched distribution
Eq.~\eqref{eq:rmt_nu0} is justified by the fact that the first
eigenvalue $\lambda_1$ is typically much smaller than the quark mass
on our lattices.

\section{Conclusions and outlook}
\label{sec:concl}

In this paper we have studied the issue of localisation of the low
Dirac eigenmodes in a toy model of QCD, consisting of unimproved
staggered fermions interacting via SU$(3)$ gauge fields on a lattice
of temporal extension $N_T=4$. While the low and high temperature
phases of QCD are connected by an analytic crossover, such a model
displays a genuine deconfining and (approximately) chirally-restoring
first-order phase transition at some critical value of the gauge
coupling $\beta$~\cite{unimproved0,unimproved1,unimproved}. This
allows us to study the relation between deconfinement, chiral
restoration and localisation in a more clear-cut setting.

Our results indicate that the onset of localisation of the lowest
Dirac eigenmodes takes place at the same critical coupling $\beta_c$ at
which the system undergoes the first-order phase transition. 
While for $\beta<\beta_c$ all the modes are delocalised, for
$\beta>\beta_c$ the lowest modes are localised up to a critical point
$\lambda=\lambda_c(\beta)$ in the spectrum, which keeps increasing as
$\beta$ is increased: this is fully analogous to what happens in
QCD. In the light of the mechanism for localisation discussed in
Ref.~\cite{GKP2}, these results support our expectation that
deconfinement triggers localisation of the lowest modes through the
appearance, in the deconfined phase, of ``islands'' in the Polyakov
line configuration, i.e., fluctuations of the Polyakov lines away from 
the ordered (trivial) value. The localised nature of these ``islands''
is also expected to play an important role in the depletion of the
spectral region around the origin~\cite{GKP2}, and therefore in the
approximate restoration of chiral symmetry. Summarising, deconfinement
appears to be the fundamental phenomenon triggering both chiral
symmetry restoration and localisation of the low modes. 

It would be interesting to study also other QCD-like models displaying
a genuine phase transition. An obvious possibility is pure
SU$(3)$ gauge theory. In contrast to that model, the one considered in
the present paper possesses a true (albeit softly broken) chiral
symmetry, and so the issue of chiral symmetry restoration is better
defined. On the other hand, the presence of a first-order phase
transition is here a lattice artifact that does not survive the
continuum limit, while it does in pure-gauge SU$(3)$. For this reason,
it would be worth studying the fate of localisation as one decreases
the lattice spacing in the latter model. 

Another interesting case would be the pure-gauge SU$(2)$ theory, where
the deconfinement transition is second-order. A study of the
near-critical behaviour of the lowest eigenmodes could then shed some
light on how (if at all) the localisation transition depends on the
order of the deconfinement transition.

\section*{Acknowledgements}
This work is partly supported by OTKA under the grant OTKA-K-113034.
TGK is supported by the Hungarian Academy of Sciences under
``Lend\"ulet'' grant No. LP2011-011.


\begin{thebibliography}{99}

\bibitem{BC} T.~Banks and A.~Casher, Nucl. Phys. B 
\nyp{169}{103}{1980}.

\bibitem{Aoki:2005vt} 
  Y.~Aoki, Z.~Fodor, S.~D.~Katz and K.~K.~Szab\'o,
  \JHEP \nyp{01}{089}{2006} 
  [hep-lat/0510084].

\bibitem{Borsanyi:2010cj} 
  S.~Bors\'anyi, G.~Endr\H odi, Z.~Fodor, A.~Jakov\'ac, S.~D.~Katz, S.~Krieg,
  C.~Ratti and K.~K.~Szab\'o,
  \JHEP \nyp{11}{077}{2010} 
  [arXiv:1007.2580 [hep-lat]].


\bibitem{GGO2}
  A.~M.~Garc\'ia-Garc\'ia and J.~C.~Osborn,
  Phys.\ Rev.\  D 
\nyp{75}{034503}{2007}
  [hep-lat/0611019].

\bibitem{KP} T.~G.~Kov\'acs and
    F.~Pittler, Phys.\ Rev.\ Lett.\ 
\nyp{105}{192001}{2010} 
  [arXiv:1006.1205 [hep-lat]].

\bibitem{BKS} 
  F.~Bruckmann, T.~G.~Kov\'acs and S.~Schierenberg,
  Phys.\ Rev.\ D 
\nyp{84}{034505}{2011} 
 [arXiv:1105.5336 [hep-lat]].


\bibitem{KP2} T.~G.~Kov\'acs and
    F.~Pittler, Phys.\ Rev.\ D 
\nyp{86}{114515}{2012}  
  [arXiv:1208.3475 [hep-lat]].






\bibitem{feri} M.~Giordano, T.~G.~Kov\'acs and
    F.~Pittler,  PoS  LATTICE 
\nyp{2013}{212}{2013}
     [arXiv:1311.1770    [hep-lat]].  

\bibitem{crit} M.~Giordano, T.~G.~Kov\'acs and F.~Pittler, 
    Phys.\ Rev.\ Lett.\ 
\nyp{112}{102002}{2014} 
  [arXiv:1312.1179 [hep-lat]].

\bibitem{KGT} T.~G.~Kov\'acs,
  Phys.\ Rev.\ Lett.\ 
\nyp{104}{031601}{2010}   
[arXiv:0906.5373 [hep-lat]]. 


\bibitem{Cossu:2016scb}
  G.~Cossu and S.~Hashimoto,
  \JHEP {\bf 06} (2016) 056 [arXiv:1604.00768 [hep-lat]].



\bibitem{GKP}
  M.~Giordano, T.~G.~Kov\'acs and F.~Pittler,
   \JHEP \nyp{04}{112}{2015}
  [arXiv:1502.02532 [hep-lat]].

\bibitem{GKP2}
  M.~Giordano, T.~G.~Kov\'acs and F.~Pittler,
   \JHEP \nyp{06}{007}{2016}
  [arXiv:1603.09548 [hep-lat]].


\bibitem{GGO} 
  A.~M.~Garc\'ia-Garc\'ia and J.~C.~Osborn,
  Nucl.\ Phys.\ A 
\nyp{770}{141}{2006} 
  [hep-lat/0512025].


\bibitem{Shuryak:1981ff}
  E.~V.~Shuryak,
  Nucl.\ Phys.\ B {\bf 203} (1982) 93.
\bibitem{Shuryak:1982dp}
  E.~V.~Shuryak,
  Nucl.\ Phys.\ B {\bf 203} (1982) 116.
\bibitem{Shuryak:1982hk}
  E.~V.~Shuryak,
  Nucl.\ Phys.\ B {\bf 203} (1982) 140.



\bibitem{Diakonov:1984vw}
  D.~Diakonov and V.~Y.~Petrov,
  Phys.\ Lett.\ B {\bf 147} (1984) 351.

\bibitem{Diakonov:1985eg}
  D.~Diakonov and V.~Y.~Petrov,
  Nucl.\ Phys.\ B {\bf 272} (1986) 457.

\bibitem{Diakonov:1995ea}
  D.~Diakonov,
  Proc.\ Int.\ Sch.\ Phys.\ Fermi {\bf 130} (1996) 397
  [hep-ph/9602375].

\bibitem{Smilga:1992yp}
  A.~V.~Smilga,
  Phys.\ Rev.\ D {\bf 46} (1992) 5598.

\bibitem{Janik:1998ki}
  R.~A.~Janik, M.~A.~Nowak, G.~Papp and I.~Zahed,
  Phys.\ Rev.\ Lett.\  {\bf 81} (1998) 264
  [hep-ph/9803289].

\bibitem{Osborn:1998nm}
  J.~C.~Osborn and J.~J.~M.~Verbaarschot,
  Phys.\ Rev.\ Lett.\  {\bf 81} (1998) 268
  [hep-ph/9807490].

\bibitem{Osborn:1998nf}
  J.~C.~Osborn and J.~J.~M.~Verbaarschot,
  Nucl.\ Phys.\ B {\bf 525} (1998) 738
  [hep-ph/9803419].

\bibitem{GarciaGarcia:2003mn}
  A.~M.~Garc\'ia-Garc\'ia and J.~C.~Osborn,
  Phys.\ Rev.\ Lett.\  {\bf 93} (2004) 132002
  [hep-th/0312146].



\bibitem{Diakonov:2009jq}
  D.~Diakonov,
  Nucl.\ Phys.\ Proc.\ Suppl.\  {\bf 195} (2009) 5
  [arXiv:0906.2456 [hep-ph]].

\bibitem{Bruckmann:2003yq}
  F.~Bruckmann, D.~N\'ogr\'adi and P.~van Baal,
  Acta Phys.\ Polon.\ B {\bf 34} (2003) 5717
  [hep-th/0309008].

\bibitem{Shuryak:2012aa}
  E.~Shuryak and T.~Sulejmanpasic,
  Phys.\ Rev.\ D {\bf 86} (2012) 036001
  [arXiv:1201.5624 [hep-ph]].

\bibitem{Poppitz:2012nz}
  E.~Poppitz, T.~Sch\"afer and M.~\"Unsal,
  JHEP {\bf 1303} (2013) 087
  [arXiv:1212.1238 [hep-th]].



\bibitem{unimproved0}
F.~Karsch, E.~Laermann and C.~Schmidt,
  Phys.\ Lett.\ B {\bf 520} (2001) 41
  [hep-lat/0107020].

\bibitem{unimproved1}
P.~de Forcrand and O.~Philipsen,
  Nucl.\ Phys.\ B {\bf 673} (2003) 170
  [hep-lat/0307020].

\bibitem{unimproved}
P.~de Forcrand and O.~Philipsen, \JHEP 
\nyp{11}{012}{2008}
[arXiv:0808.1096 [hep-lat]].

\bibitem{GKKP} M.~Giordano, T.~G.~Kov\'acs, S.~D.~Katz and F.~Pittler, 
PoS LATTICE  \nyp{2014}{214}{2014} [arXiv:1410.8392 [hep-lat]].



\bibitem{Giordano:2014qna}
  M.~Giordano, T.~G.~Kov\'acs and F.~Pittler,
   Int.\ J.\ Mod.\ Phys.\ A \nyp{29}{1445005}{2014}. 
  [arXiv:1409.5210 [hep-lat]].



\bibitem{VWrev}
  J.~J.~M.~Verbaarschot and T.~Wettig,
  Ann.\ Rev.\ Nucl.\ Part.\ Sci.\  
\nyp{50}{343}{2000} 
  [hep-ph/0003017].

\bibitem{deF}
  P.~de Forcrand,
  AIP Conf.\ Proc.\  
\nyp{892}{29}{2007} 
  [hep-lat/0611034].

\bibitem{Tomiya:2016jwr}
  A.~Tomiya, G.~Cossu, S.~Aoki, H.~Fukaya, S.~Hashimoto, T.~Kaneko and J.~Noaki,
  arXiv:1612.01908 [hep-lat].



\bibitem{Anderson58}
  P.~W.~Anderson, Phys.\ Rev.\ 
\nyp{109}{1492}{1958}.

\bibitem{LR}
  P.~A.~Lee and T.~V.~Ramakrishnan,
  Rev.\ Mod.\ Phys.\  
\nyp{57}{287}{1985}.

\bibitem{EM}
  F.~Evers and A.~D.~Mirlin,
  Rev.\ Mod.\ Phys.\  
\nyp{80}{1355}{2008}
[arXiv:0707.4378 [cond-mat.mes-hall]]. 


\bibitem{nu_unitary}
K.~Slevin and T.~Ohtsuki, Phys.\ Rev.\ Lett.\ 
\nyp{78}{4083}{1997}
[cond-mat/9704192 [cond-mat.dis-nn]].

\bibitem{Mehta}
  M.~Mehta, {\it Random Matrices}
(Academic Press, San Diego, 1991).


\bibitem{UGPKV} 
  L.~Ujfalusi, M.~Giordano, F.~Pittler, T.~G.~Kov\'acs and I.~Varga,
Phys.\ Rev.\ D \nyp{92}{094513}{2015} 
[arXiv:1507.02162 [cond-mat.dis-nn]].

\bibitem{UV}
L.~Ujfalusi and I.~Varga,
  Phys.\ Rev.\ B \nyp{91}{184206}{2015}
  [arXiv:1501.02147 [cond-mat.dis-nn]].

\bibitem{SSSLS} 
    B.~I.~Shklovskii, B.~Shapiro,  
  B.~R.~Sears, P.~Lambrianides and H.~B.~Shore, Phys. Rev. B {\bf
    47} (1993) 11487.

\bibitem{HS} E.~Hofstetter
    and M.~Schreiber, Phys. Rev. B {\bf 49} (1994) 14726.

\bibitem{Forrester} P.~J.~Forrester, Nucl.\ Phys.\ B \nyp{402}{709}{1993}.

\bibitem{Nishigaki:1998is}
  S.~M.~Nishigaki, P.~H.~Damgaard and T.~Wettig,
  Phys.\ Rev.\ D {\bf 58} (1998) 087704
  [hep-th/9803007].


\bibitem{Nishigaki:2016nka}
  S.~M.~Nishigaki,
  PoS LATTICE \nyp{2015}{057}{2016}
  [arXiv:1606.00276 [hep-lat]].


\bibitem{shape_analysis} 
I.~Varga, E.~Hofstetter, M.~Schreiber and J.~Pipek, Phys.\ Rev.\ B
\nyp{52}{7783}{1995}. 

\bibitem{Nishigaki:2013uya}
  S.~M.~Nishigaki, M.~Giordano, T.~G.~Kov\'acs and F.~Pittler,
  PoS LATTICE \nyp{2013}{018}{2014}
  [arXiv:1312.3286 [hep-lat]].


\end{thebibliography}
\end{document}